\documentclass[aps,preprint]{revtex4}%
\usepackage{amsfonts}
\usepackage{amsmath}
\usepackage{amssymb}
\usepackage{graphicx}%
\providecommand{\U}[1]{\protect\rule{.1in}{.1in}}

\begin{document}

\preprint{ }
\title{Preferred basis without decoherence}
\author{Roberto Laura}
\affiliation{Facultad de Ciencias Exactas, Ingenier\'{\i}a y
Agrimensura (UNR) and Instituto de F\'{\i}sica Rosario (CONICET -
UNR), Av. Pellegrini 250, 2000 Rosario, Argentina}
\email{rlaura@fceia.unr.edu.ar}
\author{Leonardo Vanni}
\affiliation{Instituto de Astronom\'{\i}a y F\'{\i}sica del
Espacio (UBA - CONICET). Casilla de Correos 67, Sucursal 28, 1428
Buenos Aires, Argentina.} \email{lv@iafe.uba.ar}


\begin{abstract}
The aim of this paper is to argue that the \textquotedblleft
preferred basis problem\textquotedblright\ is not a real problem
in measurement. We will show that, given an apparatus, among the
infinite corrrelations that can be established in the final state
by means of a change of basis, one and only one makes physical
sense. It is the apparatus, through its interaction Hamiltonian,
what selects a single basis and determines the observable to be
measured, even without decoherence.
\end{abstract}

\date{diciembre  2008}

\maketitle

\section{Introduction}

Let us consider the measurement of the observable $A=\sum a_{i}|a_{i}\rangle
\langle a_{i}|$ of a system $S$, by means of an apparatus $M$ with a pointer
observable $Z=\sum z_{i}|z_{i}\rangle \langle z_{i}|$. Let us suppose that $%
\{|a_{i}\rangle \}$ is a basis of the Hilbert space of $S$, and $%
\{|z_{i}\rangle \}$ is a basis of the Hilbert space of $M$. Before the
interaction between $S$ and $M$, $S$ is in the state $|\varphi \rangle =\sum
c_{i}|a_{i}\rangle $, and $M$ is prepared in a ready-to-measure state $%
|z_{0}\rangle $. According to the von Neumann model of measurement \cite%
{vonNeuman}, the interaction introduces a correlation between the
eigenstates of $A$ and the eigenstates of $Z$:%
\begin{equation}
|\varphi \rangle =\sum_{i}c_{i}|a_{i}\rangle |z_{0}\rangle \hspace{1cm}%
\longrightarrow \hspace{1cm}|\psi \rangle =U_{\Delta t}\,|\varphi \rangle
=\sum_{i}c_{i}|a_{i}\rangle |z_{i}\rangle   \label{1-1}
\end{equation}

\noindent where $U_{\Delta t}=e^{-iH_{int}\Delta t}$ is the evolution
operator corresponding to the interaction Hamiltonian $H_{int}$. After the
time $\Delta t$, the interaction ends and the correlation is established:
since then it is assumed that, if the pointer $Z$ acquires a value $z_{i}$,
then the value of $A$ measured in the process is $a_{i}$ (see \cite%
{Mittelstaedt}, \cite{Mittelstaedt Lathi}).

Eq. (\ref{1-1}) expresses the biorthonormal decomposition of the state $%
|\psi \rangle $ of the composite system $S+M$, (see \cite{Schmidt}): it
establishes the correlation that defines the measurement. Nevertheless, such
a decomposition is not always unique. \ In particular, if the coefficients $%
c_{i}$ are all equal in absolute value, it is possible to introduce a change
of basis $\{|a_{i}\rangle \}\rightarrow \{|a_{i}^{\prime }\rangle \}$ and $%
\{|z_{i}\rangle \}\rightarrow \{|z_{i}^{\prime }\rangle \}$ (see Appendix A)
such that

\begin{equation}
|\psi \rangle =\sum_{i}c_{i}|a_{i}\rangle |z_{i}\rangle
=\sum_{i}c_{i}^{\prime }|a_{i}^{\prime }\rangle |z_{i}^{\prime }\rangle
\label{1-2}
\end{equation}%
This result is what leads to the so-called \textquotedblleft preferred basis
problem\textquotedblright\ \cite{Max}: since the two different sets of
outcomes, $\{z_{i}\}$ and $\{z_{i}^{\prime }\}$, are both admissible for
same measurement, there is an intrinsic ambiguity about what observable is
measured on the system $S$. In other words, if $Z$ acquires a value $z_{i}$,
$A$ is measured according to the first expansion of $|\psi \rangle $; but if
$Z$ acquires a value $z_{i}^{\prime }$, then $A^{\prime }$ is measured
according to the second expansion. Therefore, the observer could not know
whether the measured observable is $A$ or $A^{\prime }$, even in the case
that $[A,A^{\prime }]\not=0$. \ From the viewpoint of the decoherence
program, if the composite system $S+M$ remains isolated, this problem cannot
be solved. \ For this reason, a further interaction with the environment is
necessary in order to determine the preferred $-$pointer$-$ basis: it is
such an interaction what, via decoherence, selects the observable to be
measured. This argument has been widely accepted in the literature (see, for
instance \cite{Max}).

The aim of this paper is conceptual: our purpose is to argue that the
\textquotedblleft preferred basis problem\textquotedblright\ is not a real
problem in measurement. Instead of trying to solve the supposed difficulty
by adding an interaction after the correlation given in eq. (\ref{1-2}), we
will analyze the process leading to that correlation. On this basis we will
show that, given an apparatus, among the infinite correlations that can be
established in the final state by means of a change of basis, one and only
one makes physical sense. It is the apparatus, through its interaction
Hamiltonian, what selects a single basis and determines the observable to be
measured, even when the composite system $S+M$ is a closed system.

\section{One set of outcomes}

As we have pointed out, if the coefficients $c_{i}$ that span the initial
state of the system are all equal, the change of basis is possible:%
\begin{align}
|\varphi \rangle =\sum_{i}c_{i}|a_{i}\rangle |z_{0}\rangle \hspace{1cm}%
\rightarrow \hspace{1cm}|\psi \rangle =U|\varphi \rangle &
=\sum_{i}c_{i}|a_{i}\rangle |z_{i}\rangle   \label{2-1} \\
& =\sum_{i}c_{i}^{\prime }|a_{i}^{\prime }\rangle |z_{i}^{\prime }\rangle
\label{2-2}
\end{align}%
The equality between eqs. (\ref{2-1}) and (\ref{2-2}) leads to the preferred
basis problem. However, this would really be a problem if there existed an
apparatus where the two different sets $\{z_{i}\}$ and $\{z_{i}^{\prime }\}$
could both be the outcomes of the measurement, with no modification at all
in the experimental arrangement \footnote{%
This requirement relies on the fact that any difference in the
experimental arrangement would allow us to distinguish between the
two bases and, as a consquence, to know what observable is
measured. For instance, in the Stern-Gerlach measurmeent, the
rotation of the magnetic field from direction $z$ to direction $x$
would allow us to know that $S_{z}$ was measured in the first
case, and $S_{x}$ was measured in the second case.}.  Only in this
situation the decompositions (\ref{2-1}) and (\ref{2-2}) would be
both physically meaningful, and their equality would actually
imply the ambiguity introduced by the problem.

Let us suppose that such an apparatus exists and, as a consequence, eqs. (%
\ref{2-1}) and (\ref{2-2}) really imply different sets of outcomes of the
same measuring apparatus. For simplicity, we consider an ideal measurement,
where the sets of states $\{|z_{i}\rangle \}$ and $\{|z_{i}^{\prime }\rangle
\}$ share the ready-to-measure state, $|z_{0}\rangle =|z_{0}^{\prime
}\rangle $. Then, from the calibration of the measurement \cite{Mittelstaedt}
in eq. (\ref{2-1}), when we prepare the system $S$ in the state $%
|a_{i}\rangle $, the pointer $Z$ will acquire the value $z_{i}$
corresponding to its eigenvector $|z_{i}\rangle $:%
\begin{equation}
|a_{i}\rangle |z_{0}\rangle \hspace{1cm}\rightarrow \hspace{1cm}%
|a_{i}\rangle |z_{i}\rangle .  \label{calibracion-con-z}
\end{equation}%
Nevertheless, from the calibration in eq. (\ref{2-2}), when we prepare $S$
in the state $|a_{i}^{\prime }\rangle $, $Z$ will acquire the value $%
z_{i}^{\prime }$ corresponding to $|z_{i}^{\prime }\rangle $:%
\begin{equation}
|a_{i}^{\prime }\rangle |z_{0}\rangle \hspace{1cm}\rightarrow \hspace{1cm}%
|a_{i}^{\prime }\rangle |z_{i}^{\prime }\rangle .  \label{calibracion-con-z'}
\end{equation}%
Now, since $\{|a_{i}\rangle \}$ and $\{|a_{i}^{\prime }\rangle \}$ are two
bases of the Hilbert space of the system $S$, we can introduce the change of
basis $|a_{i}^{\prime }\rangle =\sum_{k}\alpha _{k}|a_{k}\rangle $. Then, if
we introduce this change of basis in eq. (\ref{calibracion-con-z}), and
given the linearity of the evolution induced by the apparatus, we also obtain%
\begin{equation}
|a_{i}^{\prime }\rangle |z_{0}\rangle =\left( \sum_{k}\alpha
_{k}|a_{k}\rangle \right) |z_{0}\rangle \hspace{1cm}\rightarrow \hspace{1cm}%
\sum_{k}\alpha _{k}|a_{k}\rangle |z_{k}\rangle
\label{falsa-calibracion-con-z'}
\end{equation}%
Let us compare eqs. (\ref{calibracion-con-z'}) and (\ref%
{falsa-calibracion-con-z'}): eq. (\ref{calibracion-con-z'}) corresponds to
the outcome $z_{i}^{\prime }$ but, on the other hand, eq. (\ref%
{falsa-calibracion-con-z'}) says that we will obtain one of the outcomes $%
\left\{ z_{i}\right\} $. Independently of any macroscopicity condition that
allows a human being to distinguish between the set $\{z_{i}\}$ and the set $%
\{z_{i}^{\prime }\}$ \cite{Ballentine}, if these sets are different, they
represent different physical outcomes \footnote{%
Otherwise, we would not distinguish between the kets $\{|z_{i}\rangle \}$
and the kets $\{|z_{i}^{\prime }\rangle \}$ in quantum mechanics. Under the
assumption of the preferred basis problem, they are distinguished by a
non-identity change of basis.}.  In other words, eqs. (\ref{calibracion-con-z'}%
) and (\ref{falsa-calibracion-con-z'}) represent different physical
processes with different physical results, one where the pointer $Z$
acquires one of the values $\{z_{i}\}$, and the other where $Z$ acquires one
of the values $\{z_{i}^{\prime }\}$. \ However, both processes begin in the
same state $|a_{i}^{\prime }\rangle $ of the system $S$: the only difference
between them is that the state $|a_{i}^{\prime }\rangle $ is decomposed in
different bases. As a consequence, if there existed an apparatus where the
two different sets $\{z_{i}\}$ and $\{z_{i}^{\prime }\}$ could both be the
outcomes of the same measurement, such an apparatus would distinguish
between the two mathematical representations of the same initial state of $S$%
: the outcomes $\left\{ z_{i}\right\} $ would indicate the
representation of the state in the basis $\{|a_{i}\rangle \}$, and
the outcomes $\left\{ z'_{i}\right\} $ would indicate the
representation of the state in the basis $\{|a'_{i}\rangle \}$.
This means that the apparatus would distinguish mathematical
entities, and this is physically absurd. Therefore, this kind of
apparatus does not exist: any physical apparatus can only supply a
single set of outcomes.

Summing up, independently of the mathematical equality between eqs. (\ref%
{calibracion-con-z'}) and (\ref{falsa-calibracion-con-z'}), one and only one
of these correlations makes physical sense. \ Although the change of basis
is mathematically correct, the previous argument shows that, with no
modification in the apparatus, either the outcomes $\{z_{i}\}$ or the
outcomes $\{z_{i}^{\prime }\}$ can be obtained, but not both. Although the
argument dissolves the supposed ambiguity about the preferred basis, it does
not tell us yet which set is effectively obtained and, then, which the
observable measured on the system $S$ is. As we will see, this question can
be answered when measurement is considered as a process.

\section{What set of the outcomes?}

The previous section proves that any measuring apparatus, if it is a quantum
apparatus, can only activate a single set of outcomes, associated with a
certain basis. In this section we will formulate an argument for specifying
that basis.

As we have seen, only one of the correlations (\ref{2-1}) or (\ref{2-2})
makes physical sense, although the equality between them holds. This
equality only expresses a change of basis on the final state, once it has
been obtained. But endowing the mathematical change of basis with a physical
content amounts to ignoring the fact that a measurement is a \textit{quantum
process} and, as such, it is governed by the dynamical equation of the
theory. In fact, the dynamics of the process is given by the evolution
operator $U$ \ through the corresponding Hamiltonian: by beginning with an
initial state, it is $U$ what produces the correlation when applied to that
initial state. Therefore, the correlation with physical meaning is that
resulting from the application of $U$ to the initial state. \ And since the
evolution operator $U$ \ is unitary, given the initial state, the
correlation corresponding to this $U$ is unique, in spite of the possibility
of performing a mathematical change of basis on the final state.

In order to rephrase the argument in a formal way, let us consider the two
following processes:%
\begin{align}
|\varphi _{A}\rangle =\sum_{i}c_{i}|a_{i}\rangle |z_{0}\rangle \hspace{1cm}&
\longrightarrow \hspace{1cm}|\psi _{A}\rangle =U|\varphi _{A}\rangle
=\sum_{i}c_{i}|a_{i}\rangle |z_{i}\rangle   \label{3-1} \\
|\varphi _{A^{\prime }}\rangle =\sum_{i}c_{i}^{\prime
}|a_{i}^{\prime }\rangle |z'_{0}\rangle \hspace{1cm}&
\longrightarrow \hspace{1cm}|\psi _{A^{\prime }}\rangle =U^{\prime
}|\varphi _{A^{\prime }}\rangle =\sum_{i}c_{i}^{\prime
}|a_{i}^{\prime }\rangle |z_{i}^{\prime }\rangle , \label{3-2}
\end{align}%
According to the argument of the basis ambiguity, $|\psi
_{A}\rangle =|\psi _{A^{\prime }}\rangle $, but represented in
different bases, which correspond to different correlations. But\
since the correlation is established by the evolution operator,
different correlations require different evolution operators
$U\not=U^{\prime }$ and, then, different interaction Hamiltonians
$H\neq H^{\prime }$ which, in turn, correspond to
different apparatuses. This means that the arrows of eqs. (\ref{3-1}) and (%
\ref{3-2}) represent different measurement processes: with the apparatus
that induces the evolution given by $U$, we obtain an outcome belonging to
the set $\{z_{i}\}$ and never an outcome of the set $\{z_{i}^{\prime }\}$,
and analogously for the apparatus corresponding to $U^{\prime }$.

By contrast with the argument of the basis ambiguity, which only
focuses on the final state, our argument traces the process back
in order to see what dynamical evolution leads to that final
state. It is represented in the next figure:

\setlength{\unitlength}{0.05in}
\begin{picture}(15,50)(10,-25)
\put(45,15){\footnotesize $Measurement \, process\,  before\,
the\, corelation $} \thinlines \put(30,0){\line(1,0){20}}
\thinlines \put(30,-10){\line(1,0){20}} \thinlines
\put(50,0){\line(2,-1){10}} \thinlines
\put(50,-10){\line(2,1){10}} \put(22,0){$|\varphi_A \rangle $}
\put(22,-11){$|\varphi_{A^{\prime }} \rangle $} \put(35,
2){$|\psi_A \rangle=U |\varphi_A \rangle $} \put(35,
-13){$|\psi_{A^{\prime }} \rangle=U' |\varphi_{A^{\prime }}
\rangle $} \put(35, -6){$U \not =U'$} \put(65, -6){$|\psi_A
\rangle=|\psi_{A^{\prime }} \rangle=\sum_i c_i |a_i\rangle
|z_i\rangle=\sum_i c'_i |a'_i\rangle |z'_i\rangle $}
\end{picture}

 With this strategy, it is easy to show that different sets
 of outcomes finally imply different measuring apparatuses. As we can
see, if one of the outcomes $z_{i}$ is obtained, the preferred
basis is the eigenbasis of $A$, and the correlation corresponding
to $|\psi _{A}\rangle $ is introduced by $U$, that is, by the
apparatus with Hamiltonian $H$: this apparatus measures the
observable $A$ of $S$. Analogously, if one of the outcomes
$z_{i}^{\prime }$ is obtained, the preferred basis is the
eigenbasis of $A^{\prime }$, but now the correlation corresponding
to $|\psi_{A^{\prime }}\rangle $ is introduced by $U^{\prime } $,
that is, by a different apparatus with Hamiltonian $H^{\prime }$:
this second apparatus is what measures the observable $A^{\prime
}$ of $S$.

Summing up, the supposed basis ambiguity vanishes once one acknowledges that
a quantum measurement is not fully specified by the correlation expressed in
the final state of the system $S+M$, but is defined by the process that,
starting from the initial state, introduces the correlation in that final
state. And decoherence has played no role in this argument.

\section{Measurement of the spin}

\subsection{Stern-Gerlach measurement}

Let us consider a Stern-Gerlach device with a magnetic field in direction $z$%
, which correlates the projection $S_{z}$ of the spin of the particle with
its momentum $p_{z}$. The interaction Hamiltonian of this apparatus, after a
$\Delta t$, is given by $H_{int}=-c\,z\otimes S_{z}$ \cite{Ballentine},
where $c$ is a constant related to the magnetic momentum of the particle.
The evolution operator is, then, $U=e^{ic\,z\otimes S_{z}}$. Let us suppose
that the initial state of the composite system before the interaction is $%
|\varphi _{A}\rangle =\frac{1}{\sqrt{2}}\left( |+\rangle +|-\rangle \right)
|p_{0}\rangle $, where $|p_{0}\rangle $ is the momentum state with $p_{z}=0$%
. In this case, the supposed basis ambiguity is expressed as%
\begin{align}
|\varphi _{A}\rangle =\frac{1}{\sqrt{2}}\left( |+\rangle
+|-\rangle \right) |p_{0}\rangle \hspace{0.5 cm}\longrightarrow
\hspace{0.5cm}|\psi _{A}\rangle
=U|\varphi _{A}\rangle & =\frac{1}{\sqrt{2}}|+\rangle |p_{+}\rangle +\frac{1%
}{\sqrt{2}}|-\rangle |p_{-}\rangle  \label{evolucion-Sz} \\
& =\frac{1}{\sqrt{2}}|\odot \rangle |p_{\odot }\rangle +\frac{1}{\sqrt{2}}%
|\otimes \rangle |p_{\otimes }\rangle  \label{falsa-evolucion-Sx}
\end{align}%
where%
\begin{equation}
|\odot \rangle =\frac{1}{\sqrt{2}}\left( |+\rangle +|-\rangle \right) ,%
\hspace{1cm}|\otimes \rangle =\frac{1}{\sqrt{2}}\left( |+\rangle -|-\rangle
\right) ,  \label{alfa}
\end{equation}%
are the eigenstates of $S_{x}$, and $|p_{\odot }\rangle $, $|p_{\otimes
}\rangle $ are given by the combinations%
\begin{equation}
|p_{\odot }\rangle =\frac{1}{\sqrt{2}}\left( |p_{+}\rangle +|p_{-}\rangle
\right) ,\hspace{1cm}|p_{\otimes }\rangle =\frac{1}{\sqrt{2}}\left(
|p_{+}\rangle -|p_{-}\rangle \right)  \label{alfa-1}
\end{equation}

The evolution operator $U=e^{ic\,z\otimes S_{z}}$ applied to the initial
state $|\varphi _{A}\rangle $ produces the correlation given by eq. (\ref%
{evolucion-Sz}), and not that given by eq. (\ref{falsa-evolucion-Sx}). Then,
the outcomes that may be obtained as a result of the measurement are $p_{+}$
or $p_{-}$, and not $p_{\odot }$ or $p_{\otimes }$. In spite of the equality
between the eqs. (\ref{evolucion-Sz}) and (\ref{falsa-evolucion-Sx}), only
the correlation (\ref{evolucion-Sz}) has physical meaning, because it is
produced by $U$: there is no ambiguity about whether the apparatus measures $%
S_{z}$ or $S_{x}$. In other words, although the equality is valid, eq. (\ref%
{falsa-evolucion-Sx}) does not imply that $S_{x}$ has been measured. If we
want to measure $S_{x}$, we have to change the interaction introduced by the
apparatus by rotating the magnetic field from direction $z$ to direction $x$%
. But this amounts to change the apparatus itself, with its evolution
operator: in this new case, $U^{\prime }=e^{ic\,x\otimes S_{x}}$.

This example has an additional particular feature that allows us to discard $%
p_{\odot }$ and $p_{\otimes }$ as possible outcomes of the measurement. Even
under the ambiguity of the basis change, the states $|p_{\odot }\rangle $
and $|p_{\otimes }\rangle $, resulting from the combinations of the states $%
|p_{+}\rangle $ and $|p_{-}\rangle $, are not possible states of the
apparatus that measures $S_{x}$, as eq. (\ref{falsa-evolucion-Sx}) might
suggest. In fact, the values $p_{+}$ and $p_{-}$, corresponding to the
states $|p_{+}\rangle $ and $|p_{-}\rangle $, are associated to wavepackets
leading to \textquotedblleft spot up\textquotedblright\ and
\textquotedblleft spot down\textquotedblright\ on the screen; they
physically mean that $S_{z}$ was measured with the value $1/2$ or $-1/2$,
respectively. But, in this case, the values $p_{\odot }$ and $p_{\otimes }$,
corresponding to the states $|p_{\odot }\rangle $ and $|p_{\otimes }\rangle $%
, cannot be associated to \textquotedblleft spot right\textquotedblright ,
\textquotedblleft spot left\textquotedblright , that is, to the result that
the measurement of $S_{x}$ would produce: $|p_{\odot }\rangle $ and $%
|p_{\otimes }\rangle $ are linear combinations of $|p_{+}\rangle $ and $%
|p_{-}\rangle $, and none combination of wavepackets \textquotedblleft
up\textquotedblright\ and \textquotedblleft down\textquotedblright\ can
produce \textquotedblleft right\textquotedblright\ and \textquotedblleft
left\textquotedblright\ as a result.

\subsection{Bit-by-bit measurement.}

Let us now consider a system of spin $1/2$, for instance, a
particle (the measured system $S$), measured by another system of
spin $1/2$, for instance, an atom with two levels (the measuring
apparatus $M$). The basis of the Hilbert space of the particle is
$\{|+\rangle ,|-\rangle \}$, and the basis of the Hilbert space of
the atom is $\{|\uparrow \rangle ,|\downarrow \rangle \}$. The
interaction Hamiltonian after a $\Delta t$ is $H_{int}=-\pi
|-\rangle \langle -|\otimes |\leftarrow \rangle \langle \leftarrow
|$ (see \cite{Peres}, \cite{Paz-Zurek}). The evolution operator
is, then, $U=e^{i\pi |-\rangle \langle -|\otimes |\leftarrow
\rangle \langle \leftarrow |}$. Let us suppose that the initial
state of the composite system before the interaction is $|\varphi
_{A}\rangle =\frac{1}{\sqrt{2}}\left( |+\rangle +|-\rangle \right)
|\uparrow \rangle $, where $|\uparrow \rangle $ is the initial
state of the atom. In this case, the supposed basis ambiguity is
expressed as%
\begin{align}
|\varphi _{A}\rangle =\frac{1}{\sqrt{2}}\left( |+\rangle
+|-\rangle \right) |\uparrow \rangle \hspace{0.5cm}\longrightarrow
\hspace{0.5cm}|\psi _{A}\rangle =U|\varphi _{A}\rangle &
=\frac{1}{\sqrt{2}}|+\rangle |\uparrow \rangle
+\frac{1}{\sqrt{2}}|-\rangle |\downarrow \rangle
\label{evolucion Sz} \\
& =\frac{1}{\sqrt{2}}|\odot \rangle |\rightarrow \rangle +\frac{1}{\sqrt{2}}%
|\otimes \rangle |\leftarrow \rangle   \label{falsa evolucion Sx}
\end{align}%
where%
\begin{equation}
|\odot \rangle =\frac{1}{\sqrt{2}}\left( |+\rangle +|-\rangle \right) ,%
\hspace{1cm}|\otimes \rangle =\frac{1}{\sqrt{2}}\left( |+\rangle -|-\rangle
\right)   \label{gamma}
\end{equation}%
are the eigenstates of $S_{x}$ of the particle, and
\begin{equation}
|\rightarrow \rangle =\frac{1}{\sqrt{2}}\left( |\uparrow \rangle
+|\downarrow \rangle \right) ,\hspace{1cm}|\leftarrow \rangle =\frac{1}{%
\sqrt{2}}\left( |\uparrow \rangle -|\downarrow \rangle \right) .
\label{gamma-1}
\end{equation}%
are the eigenstates of $S_{x}$ of the atom. By contrast with the
Stern-Gerlach measurement, in this case the states $|\rightarrow \rangle $
and $|\leftarrow \rangle $, resulting from the combinations of the states $%
|\uparrow \rangle $ and $|\downarrow \rangle $, are possible states of the
apparatus that measures $S_{x}$ in the correlation (\ref{falsa evolucion Sx}%
). Nevertheless, as we have already pointed out, only one of the
correlations makes physical sense, and only one set of outcomes is
activated.  In fact, the evolution operator $U=e^{i\pi |-\rangle
\langle -|\otimes |\leftarrow \rangle \langle \leftarrow |}$
applied to the initial state $|\varphi _{A}\rangle$ produces the correlation given by eq. (\ref%
{evolucion Sz}), and not that given by (\ref{falsa evolucion Sx}).
Therefore, the outcomes that may be obtained as a result of the
measurement are $|\uparrow \rangle$ and $|\downarrow \rangle$, and
not $|\rightarrow \rangle$ and $|\leftarrow \rangle $: this fact
guarantees that the observable $S_{z}$ of the particle is
measured.

Again, if we want to measure $S_{x}$, we have to change the
apparatus: the new apparatus has to produce the correlation given
by eq. (\ref{falsa evolucion Sx}), which can be achieved if it
induce an  evolution  described by the operator $U'=e^{i\pi
|\otimes\rangle \langle \otimes|\otimes |\downarrow \rangle
\langle \downarrow |}$.

\section{Conclusions}

In his 1981 seminal work, Zurek speaks for the first time about the
\textquotedblleft \textit{ambiguity in the choice of the preferred apparatus
basis}\textquotedblright\ in measurement (see \cite{Zurek81}, p. 1516).
According to him, it is the decoherence resulting from the interaction
between the apparatus and the environment what determines \textquotedblleft
\textit{in what mixture the wave function appears to have collapsed}%
\textquotedblright\ (\cite{Zurek81}, p. 1517). This argument has been
repeatedly appealed to through the years (see \cite{Zurek82}, \cite{Zurek91}%
), and reappears in Zurek's more recent works (\cite{Paz-Zurek}, \cite%
{Zurek-largo}). Moreover, the \textquotedblleft problem of the preferred
basis\textquotedblright\ has been considered (with the \textquotedblleft
definite outcomes problem\textquotedblright ) as one of the two central
problems of quantum measurement. In his review paper on decoherence,
Schlosshauer considers that, although in the literature the definite
outcomes difficulty is typically referred to as `the measurement problem',
\textquotedblleft \textit{the preferred-basis problem is at least equally
important, since it does not make sense even to inquire about specific
outcomes if the set of possible outcomes is not clearly defined}%
\textquotedblright\ (see \cite{Max}, p. 1270). On the basis of this
assumption, the solution to preferred basis problem supplied by the
decoherence program is usually viewed as one of its main theoretical
advantages.

In this paper we have argued that this uncritically accepted problem is not
a legitimate difficulty, but a pseudo-problem. The supposed basis ambiguity
is the result of considering only a formal property of state vectors, and of
forgetting the physical process of measurement.\ In fact, when the
measurement is correctly understood as a physical process obeying the
dynamical law of quantum mechanics, the supposed ambiguity vanishes since
the preferred basis is determined by the measuring apparatus. Although the
argumentation leading to this conclusion was simple and straightforward, it
deserves to be seriously taken into account: its conceptual relevance
derives from the widespread acceptation of the basis ambiguity as a
legitimate problem in measurement.

\appendix

\section{Non-uniqueness of the biorthonormal decomposition}

Let us consider the final correlated state of the measurement:
\begin{equation}
|\psi _{A}\rangle =\sum_{i}c_{i}|a_{i}\rangle |z_{i}\rangle
\label{A-1}
\end{equation}%
If we introduce the change of basis $|a_{i}\rangle
=\sum_{j}M_{ji}|a_{j}^{\prime }\rangle $, then $|\psi _{A}\rangle $ reads
\begin{align}
|\psi _{A}\rangle & =\sum_{i}c_{i}\sum_{j}M_{ji}|a_{j}^{\prime }\rangle
|z_{i}\rangle  \nonumber \\
& =\sum_{j}|a_{j}^{\prime }\rangle
\sum_{i}c_{i}M_{ji}|z_{i}\rangle
\nonumber  \\
& =\sum_{j}c_{j}^{\prime }|a_{j}^{\prime }\rangle |z_{i}^{\prime
}\rangle \nonumber
\end{align}%
The last equality holds only when
\begin{equation}
c_{j}^{\prime }|z_{i}^{\prime }\rangle =\sum_{i}c_{i}M_{ji}|z_{i}\rangle
\label{A-5}
\end{equation}
If we multiply eq. (\ref{A-5}) by ${c_{l}^{\prime }}^{\ast }\langle
z_{l}^{\prime }|=\sum_{m}c_{m}^{\ast }M_{lm}^{\ast }\langle z_{m}|$, we
obtain
\begin{align}
{c_{l}^{\prime }}^{\ast }c_{j}^{\prime }\langle z_{l}^{\prime
}|z_{i}^{\prime }\rangle & =\sum_{m}c_{m}^{\ast }M_{lm}^{\ast }\langle
z_{m}|\sum_{i}c_{i}M_{ji}|z_{i}\rangle  \nonumber \\
{c_{l}^{\prime }}^{\ast }c_{j}^{\prime }\delta _{lj}&
=\sum_{i}|c_{i}|^{2}M_{li}^{\ast }M_{ji}  \nonumber \\
|c_{l}^{\prime }|^{2}\delta _{jl}&
=\sum_{i}|c_{i}|^{2}M_{ji}M_{il}^{\dagger },  \nonumber
\end{align}%
This shows that decomposition is not unique if, $\forall i$, $%
|c_{i}|=C=|c_{i}^{\prime }|$. In this case, $|z_{i}\rangle
=\sum_{j}M_{ij}^{\dagger }|z_{j}^{\prime }\rangle $.

\end{document}